\documentclass{article}

\usepackage[preprint]{neurips_2023}
\usepackage{tabularx}
\usepackage{booktabs}
\usepackage{rotating}

\usepackage[utf8]{inputenc} 
\usepackage[T1]{fontenc}    
\usepackage{hyperref}       
\usepackage{url}            
\usepackage{booktabs}       
\usepackage{amsfonts}       
\usepackage{nicefrac}       
\usepackage{microtype}      
\usepackage{xcolor}         

\title{Clustering Analysis of US COVID-19 Rates, Vaccine Participation, and Socioeconomic Factors}

\author{%
  Morteza Maleki \\   
  College of Computing\\
  Georgia Institute of Technology\\
  Atlanta, Georgia\\
  mmaleki3@gatech.edu
}

\begin{document}

\maketitle

\begin{abstract}
The COVID-19 pandemic has presented unprecedented challenges worldwide, with its impact varying significantly across different geographic and socioeconomic contexts. This study employs a clustering analysis to examine the diversity of responses to the pandemic within the United States, aiming to provide nuanced insights into the effectiveness of various strategies. We utilize an unsupervised machine learning approach, specifically K-Means clustering, to analyze county-level data that includes variables such as infection rates, death rates, demographic profiles, and socio-economic factors. Our analysis identifies distinct clusters of counties based on their pandemic responses and outcomes, facilitating a detailed examination of "high-performing" and "lower-performing" groups. These classifications are informed by a combination of COVID-specific datasets and broader socio-economic data, allowing for a comprehensive understanding of the factors that contribute to differing levels of pandemic impact. The findings underscore the importance of tailored public health responses that consider local conditions and capabilities. Additionally, this study introduces an innovative visualization tool that aids in hypothesis testing and further research, enhancing the ability of policymakers and public health officials to deploy more effective and targeted interventions in future health crises.
\end{abstract}

\section{Introduction}

The COVID-19 pandemic has indelibly marked global public health, economic stability, and societal norms. With millions of confirmed cases and widespread impact, the virus's rapid spread has prompted varied responses from different jurisdictions, influenced by geographic, demographic, and socio-economic factors. Initial measures in the United States focused predominantly on curtailing transmission through widespread lockdowns and mask mandates, often without consideration for regional disparities in healthcare access, political alignment, and public compliance levels.

This variability in pandemic responses and their outcomes presents a critical opportunity for analysis. Understanding the effectiveness of different strategies across diverse contexts is vital for preparing more resilient public health responses in the future. Thus, our study focuses on a data-driven approach to dissect these varied responses within the United States at the county level. 

Employing clustering analysis, this research identifies patterns and correlations between pandemic outcomes and the socio-economic characteristics of counties. By applying unsupervised machine learning techniques, specifically K-Means clustering, we categorize counties into distinct groups or clusters based on their performance in managing COVID-19—assessed through metrics such as infection rates, mortality rates, and vaccine uptake., \cite{maleki2023covid}

The primary objective of our analysis is to determine which factors contribute to a county being classified as "high-performing" or "low-performing" in its pandemic response. This categorization not only highlights effective strategies but also pinpoints areas where improvements are necessary, thereby offering targeted insights for policymakers and health officials. Moreover, the study introduces an innovative visualization tool that supports the dynamic exploration of data, enabling researchers and decision-makers to formulate and test hypotheses based on real-world outcomes.

As the global community continues to navigate the challenges of COVID-19 and future pandemics, the insights derived from this analysis aim to contribute to a more informed, agile, and region-specific response strategy that can be adapted to the unique needs of diverse populations.

The COVID-19 pandemic has impacted people worldwide, with millions of cases confirmed \cite{asita2020covid}. Initial policies in the United States focused on reducing transmission of the virus \cite{holtz2020interdependence, maleki2024clinical}, in order to improve overall public health, but may not have accounted for geographic and socioeconomic factors. This project takes a data-driven approach to COVID-19 in the United States to try to supply more nuanced recommendations on how best to manage COVID-19, including an innovative visualization to guide researchers in formulating and testing hypotheses.\cite{maleki2022social}\\

COVID-19 has had devastating impacts on the United States (US) population. The pandemic impacted certain populations differently than others based on demographics \cite{JHopkinsRelationship, kelly2021predictors, rentsch2020covid, dixon2021synchronicity}, financial status \cite{JHopkinsRelationship, falato2021financial}, behavioral / psychographics \cite{holtz2020interdependence, viswanath2021individual, liu2021hesitancy, kelly2021predictors, wanberg2020socioeconomic} and geographies \cite{holtz2020interdependence, dixon2021synchronicity}. Additionally, socio-economic factors like exposure to media and political party affiliation \cite{painter2021political} played an important role in influencing vaccine uptake, \cite{viswanath2021individual} and different jurisdictions (e.g., states, counties, federal) had varying responses on how to control and reduce risks of the pandemic (e.g., mask mandates, school closures) \cite{holtz2020interdependence, abedi2021racial}. Our objective is to identify "high-performing" groups of counties (i.e., counties that experienced below-average COVID impact), execute summary statistics about why the groups might be “high-performing”, and analyze “lower-performing” (i.e. counties that experience high COVID impact) groups of counties to suggest improvement opportunities. With this analysis, counties can use data-driven insights to instill new practices for managing COVID-19.\\

\section{Literature Survey}

Currently, the United States is following a broad, conservative approach to contain the pandemic that focuses strongly on preventing transmission \cite{holtz2020interdependence}. This is logical given the contagious nature of the virus but there might be underlying systemic differences putting certain groups of society at a bigger disadvantage than others \cite{abedi2021racial, dixon2021synchronicity}. Hesitancy to vaccines has been cited as a barrier to effective control of COVID-19,6 but we need to understand the root-cause of this hesitancy to better handle future unforeseen circumstances. Previous analysis on the spread of the pandemic that is geography-specific is mostly limited to the number of “cases”, “deaths” and “vaccines” \cite{vadyala2021prediction, hutagalung2021covid, hippisley2021risk, zubair2020efficient}. This is due to the availability of data broadly \cite{mcdonald2021can} and the limitations of those data that do exist \cite{dixon2021synchronicity, marivate2020use, miller2022effects}. While current approaches are informative in nature, they lack the prescriptive aspect of data analysis and fail to provide recommendations as to what actions, if any, should be taken to promote vaccination.\\

In our approach, an unsupervised learning analysis of COVID-19 infection, death rate, and many other population factors by county was carried out to identify factors that led to the specific COVID-19 outcomes within counties. Counties are clustered into like-groupings (e.g., k-clusters), and the groupings are analyzed for key summary statistics like average hospitalization rate, average vaccine adoption, and average education status. By doing so, assumptions can then be drawn from the summary statistics as to what attributes to a “higher performing” county vs. a “lower performing” county. The groupings can then be visually analyzed by utilizing the first two principal components to identify the separation and distance from other groupings.\\

While K-Means has been used in some existing studies, this approach combines the algorithm with the demographic and socio-economic data \cite{painter2021political, JHopkinsRelationship, viswanath2021individual, abedi2021racial, liu2021hesitancy, kelly2021predictors, rentsch2020covid, dixon2021synchronicity} by county and provides a fresh perspective on the drivers of pandemic spread to identify communities that are likely to respond well and poorly to unforeseen future events \cite{falato2021financial} and help identify pain points that need to be worked on at the policy level to help communities be better prepared for a similar health emergency in the future.\\

Recent studies have increasingly employed clustering algorithms to explore various dimensions of the COVID-19 pandemic. In particular, machine learning techniques have been pivotal in analyzing the geographic and temporal dynamics of the virus’s spread. For instance, Kriegel et al. (2021) utilized advanced clustering techniques to examine mobility data and its correlation with virus transmission rates across Europe, highlighting how mobility patterns could predict outbreak severity \cite{kriegel2021}.

The integration of socioeconomic data into clustering analyses has also been a critical development, revealing disparities in health outcomes. A study by Smith et al. (2022) combined health data with economic indicators using a multi-layered clustering approach. This study identified high-risk areas and suggested targeted interventions, demonstrating the complex interplay between economic conditions and health vulnerabilities during the pandemic \cite{smith2022}.

On a global scale, comparative studies using clustering analysis have shed light on the effectiveness of various national policies and their outcomes. A notable contribution by Zhang et al. (2023) compared pandemic management strategies in over thirty countries, using unsupervised clustering to classify countries based on policy effectiveness and public health outcomes. Their findings emphasized the role of early intervention and robust health infrastructure in mitigating the pandemic's impact \cite{zhang2023}.

These studies underscore the utility of clustering analysis in understanding and managing pandemics by linking data-driven insights with practical policy applications. They also highlight the potential for these techniques to facilitate better preparedness and response strategies for future global health crises.

\section{Materials and Methods}

This section outlines the structured approach undertaken to gather, clean, analyze, and visualize the data used in our clustering analysis of COVID-19 responses at the county level in the United States. The methodology is divided into three main areas: Data Collection and Cleaning, Computation and Analysis, and Visualization.

\subsection{Data Collection and Cleaning}

The first step in the process was a literature review to identify variables relevant to our analysis, followed by a compilation of relevant, available datasets. Socio-economic datasets that report on education level, election decisions, crime rate, unemployment rate, mask usage and more were combined with COVID-specific datasets (COVID tests conducted, confirmed cases, vaccination rates, COVID-related deaths) and the capacity to handle covid cases (testing capacity, no of testing clinics, hospital bed capacity, percentage of essential workers). The datasets were chosen such that they were able to be extrapolated to a United States county level. The data is publicly available and sourced from Economic Research Service at USDA \cite{ERSdatasource}, The US COVID Atlas \cite{COVIDdatasource}, Center for Disease Control and Prevention \cite{CDCdatasource}, and Census Bureau \cite{CovidCensusdatasource}.\\

One of the challenges in this step was to aggregate, clean and transform 18 datasets spanning across 3k+ counties into one master dataset. Given the unavailability of county-level datasets for federal education investment, hospital bed occupancy and covid-testing facility, we utilized datasets that were segmented by latitudes/longitudes or business addresses and then used Google Cloud provided APIs to derive corresponding zip codes and aggregate into county-level data for use in the master repository. We removed variables that had more than 50\% null values (ICU bed occupancy rate, number of covid testing clinics and total federal education investment), as well as any duplicates. Python code was developed for all the steps mentioned above.\\

The first step involved conducting a comprehensive literature review to identify relevant variables that influence COVID-19 outcomes. Data sources were meticulously selected to ensure a robust and comprehensive dataset covering socio-economic factors and COVID-19 specific metrics. Data was sourced from various reputable agencies including the Economic Research Service at the USDA, the US COVID Atlas, the Centers for Disease Control and Prevention, and the U.S. Census Bureau.

Given the diverse and extensive datasets, considerable effort was dedicated to data cleaning and integration. This involved aggregating data from over 3,000 U.S. counties, normalizing various metrics to a common scale, and handling missing or incomplete data entries. Variables with more than 50\% missing values were excluded to maintain the integrity of the analysis. The cleaned datasets were then consolidated into a master dataset that served as the foundation for subsequent analytical processes.

\subsection{Computation and Analysis}

The core of our computational approach involved the use of unsupervised machine learning techniques to cluster county data based on their pandemic response efficacy. The K-Means clustering algorithm was employed, which is well-suited for partitioning large datasets into clusters that minimize the variance within each cluster.

Prior to clustering, the data underwent several preprocessing steps:
\begin{itemize}
    \item \textbf{Normalization:} Each numeric feature was normalized to zero mean and unit variance to ensure equal weighting during the clustering process.
    \item \textbf{Dimensionality Reduction:} Principal Component Analysis (PCA) was applied to reduce the dimensionality of the data while retaining the most informative features. This step was crucial for enhancing the computational efficiency and interpretability of the clustering results.
\end{itemize}

Different numbers of clusters (k-values) were tested, ranging from 2 to 20, to determine the optimal clustering solution using the elbow method. This method assesses the percentage of variance explained as a function of the number of clusters, and it helped identify a clear point where the addition of another cluster does not give a significant improvement in variance explained.

\subsection{Visualization}

The final step involved the development of an interactive visualization tool using Tableau. This tool allows users to explore the clustered data through various lenses by adjusting parameters and selecting different socio-economic and health-related variables. Interactive maps and charts provide a user-friendly interface for visualizing complex datasets and facilitate the formulation and testing of hypotheses regarding the factors influencing COVID-19 outcomes.

This dynamic visualization supports a deeper engagement with the data, enabling stakeholders, researchers, and policymakers to gain actionable insights into the geographic and demographic factors affecting pandemic responses and outcomes.


\section{Results}

As previously discussed, we ran experiments to test varying values of $K$ when utilizing the unsupervised K-Means clustering algorithm. From evaluating the Inertia and Silhouette plots, the optimal number of clusters appeared to be greater than or equal to 2 and less than or equal to 5. To further evaluate the optimal number of clusters, we reviewed summary statistics for each variable for each cluster in the various cluster sets (i.e., 2 clusters, 3 clusters, 4 clusters, 5 clusters). When evaluating 5 clusters, the summary statistics of each cluster did not differ much cluster to cluster, leaving us with a vague representation of the various counties. When evaluating 2 clusters, the summary statistics of each cluster were greatly different, but we did not like the binary representations of counties (e.g., good performance vs. bad performance). Therefore, we evaluated 3 and 4 clusters, and found that the 3 clusters provided an effective decision boundary with clear segmentation between the clusters that could also be visualized in proceeding analysis (e.g., the interactive map).

\begin{figure*}
\centering
\includegraphics[width=15.5cm]{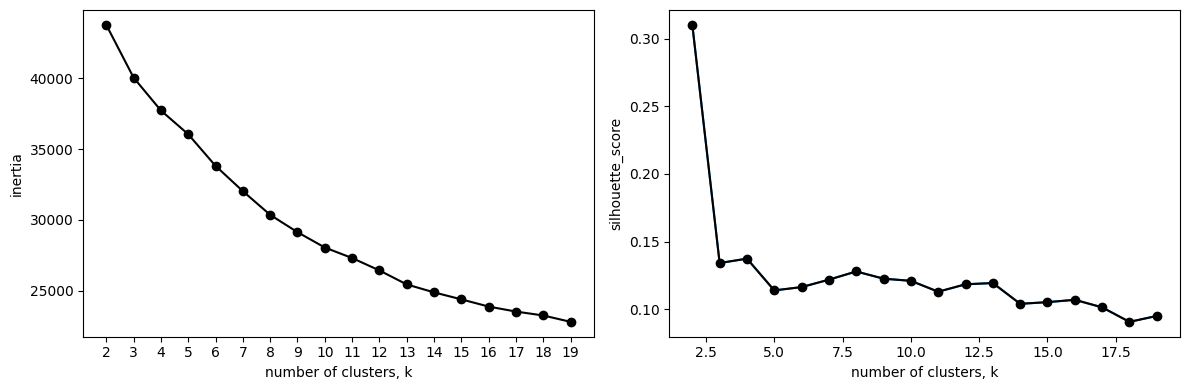}
\caption{Inertia and Silhouette curves to determine the optimal number of k clusters for K-Means Algorithm
}
\end{figure*} 

\begin{figure*}
\centering
\includegraphics[width=10cm]{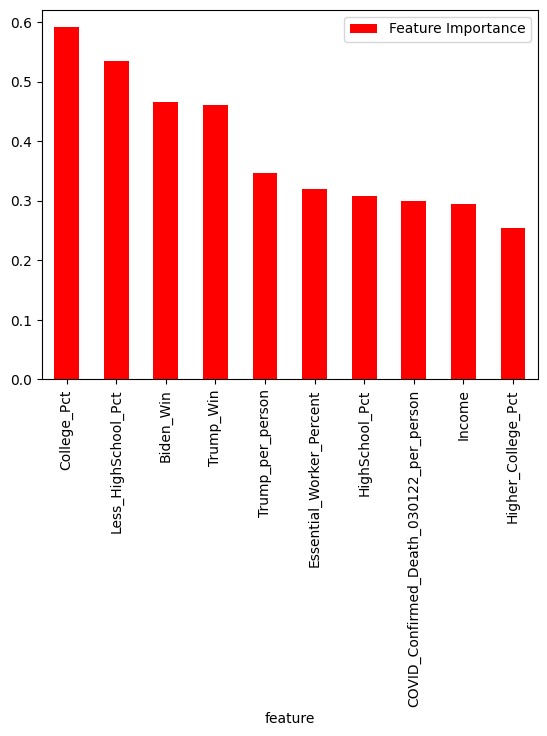}
\caption{Feature Importance Results – Top 10 Features in Minimizing Distance Between Cluster Centroids and Clustered Counties}
\end{figure*} 

\begin{figure*}
\centering
\includegraphics[width=15.5cm]{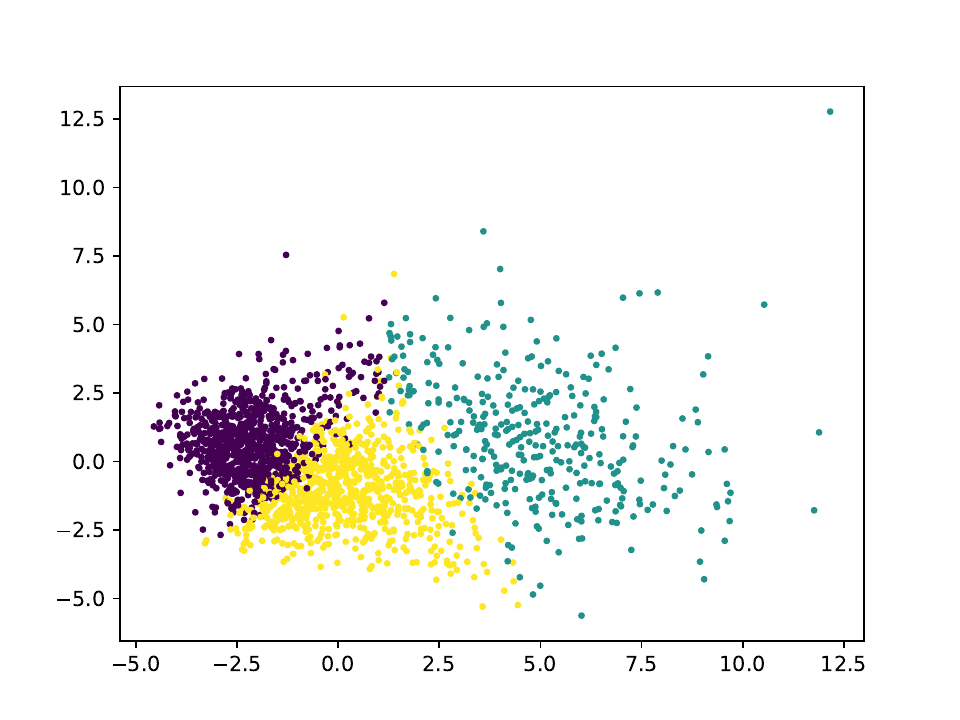}
\caption{2 dimensional visuazliation of Principal Component Analysis (PCA) on the features of the collected datasets}
\end{figure*}

We studied the characteristics of each cluster using Principal Component Analysis. Exploratory graphs that can be seen in Figure 2, namely biplot and radar chart, helped visualize, understand, and summarize the dominating variables in each cluster.

Based on the above, we arrived at the following results that can be observed in Table 1. This was followed by a cluster analysis between similar communities across state lines, and then re-focused the analysis to state-specific results. Our main findings here are:
\begin{itemize}
    \item The mask usage score was highest for cluster 1 which is our high performing cluster and lowest for Cluster 0 which is our low performing cluster. Mask usage has been encouraged as a direct means to contain the virus and our results corroborate it so far - Cluster 1 had the least COVID positivity rate and confirmed COVID cases per person.
    \item Cluster 2 (Medium performer counties) is associated with higher levels of Biden votership and higher population density – provide incremental information about the results.
    \item The state of Georgia had more than 60\% of counties included in the analysis in ‘Cluster 0’, while the state of New Jersey (several of the group members’ home state) had more than 60\% of counties included in the analysis in ‘Cluster 1’.
\end{itemize}

\begin{figure*}
\centering
\includegraphics[width=15.5cm]{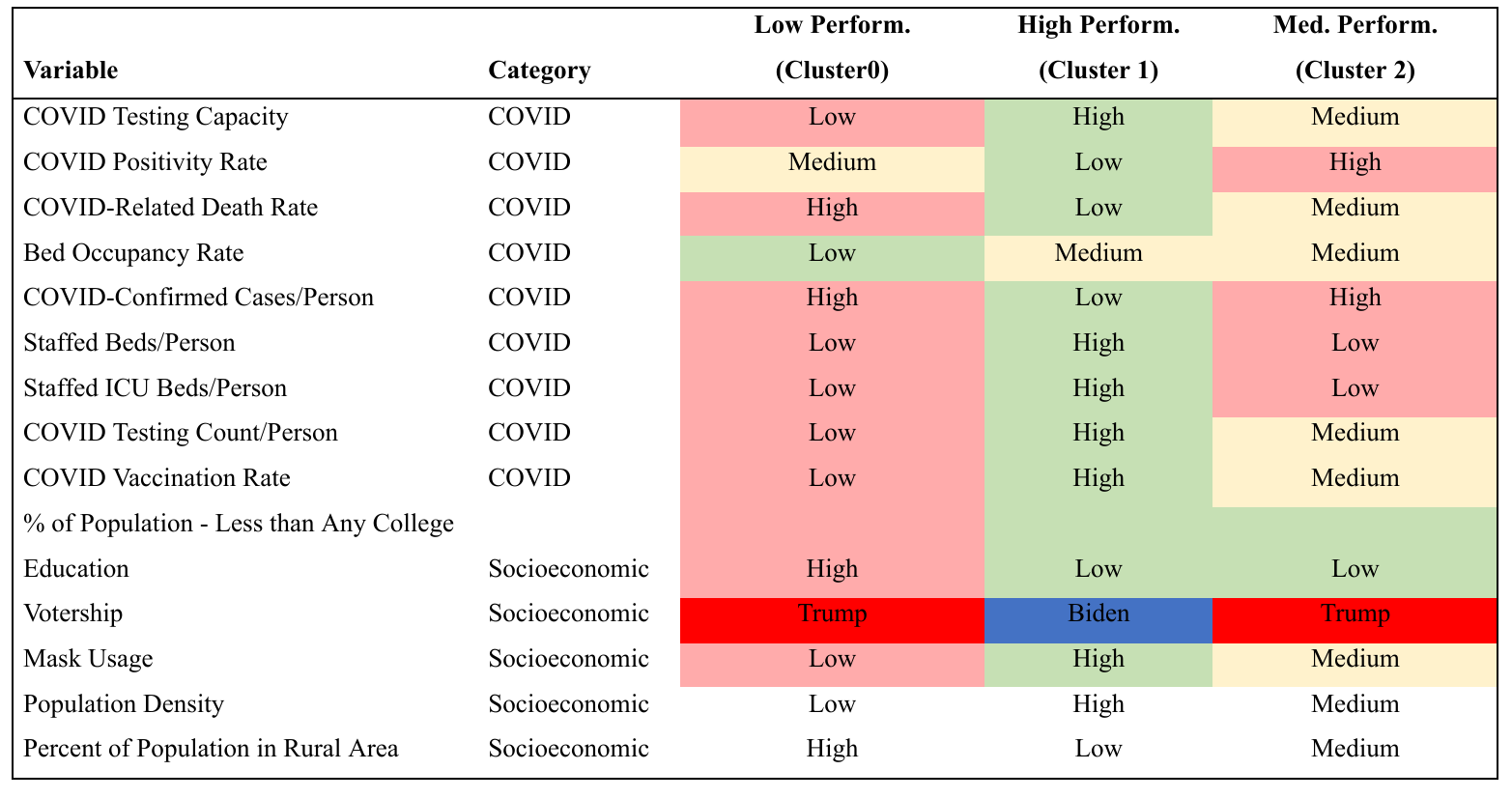}
\caption{Variable insights for each cluster in K-Means cluster analysis; High, Medium and Low Rating is relative to the clusters in the dataset  }
\end{figure*} 

In addition to these qualitative categorizations of each feature across the three clusters, the group employed a novel feature importance approach to k-means to better understand the impact of various included county characteristics. While k-means is an unsupervised learning algorithm, it is possible to develop a sense of each feature’s impact on cluster determination by delving into the construction of each cluster. Specifically, the group used a publicly available approach to interpreting k-means. The k-means algorithm is based on minimizing the distance between points grouped within each cluster (the “Within-Cluster Sum of Squares”), and this interpretation approach identifies which specific features were responsible for the most minimization of distance between each clustered county and the cluster centroid. The team ran this analysis on the generated k-means clusters; the below chart shows the derived ‘feature importance’ of the top 10 features determined by this approach.

As shown, a mix of socioeconomic (education level, political affiliation) and COVID-specific factors (complete vaccination rates) were among the features that most impacted cluster determination. This experimental approach to ‘feature importance’ helps affirm the importance of combining the multifaceted datasets used in the project to cluster COVID impact across U.S. counties.

Using Tableau, an interactive U.S. map was created to visualize the data by county. High, medium, and low performing counties are displayed via color scale. The visualization includes a menu for the user to select from a list of socio-economic factors that have been identified by the team as influencers of COVID-19 outcomes. Once a variable is selected, the pertinent value for the corresponding county can be visualized on the map; more display options can be incorporated by updating an array that defines the menu configuration. A tooltip displays the data for each county upon mouseover.

\begin{figure*}
\centering
\includegraphics[width=15.5cm]{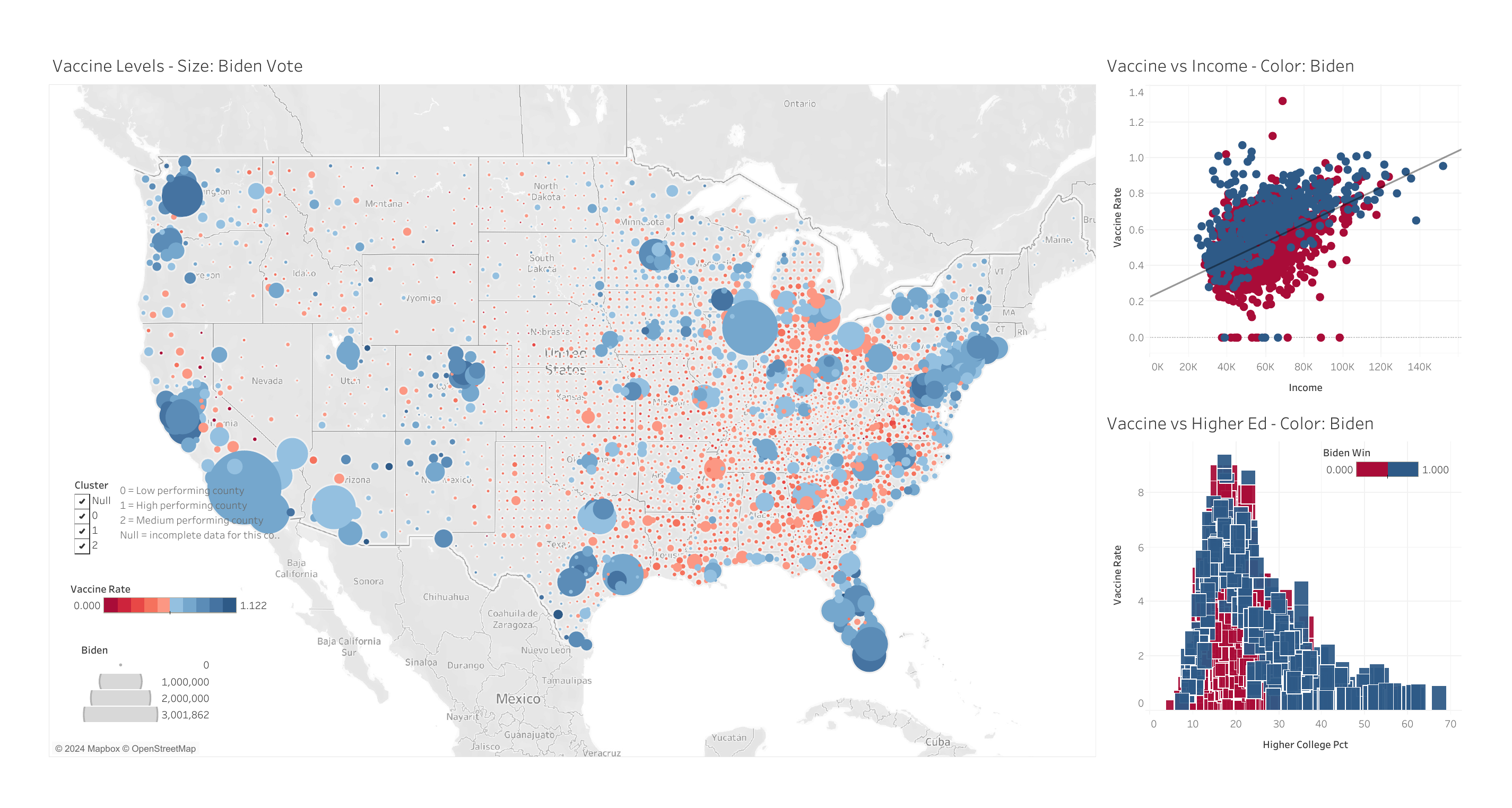}
\caption{Map Visualization to interact with results from clustering analysis}
\end{figure*}

Utilizing an interactive map visual like the one shown in Figure 4 allows users to explore the data in new and meaningful ways, such as learning from others, identifying large scale patterns, and drilling down into hypothesis testing. An example of “learning from others” can be if you are a “low performing county”, you can compare yourself to nearby “high performing counties” and learn from what the high performer is doing differently (e.g., a nearby higher performing county might have higher vaccine rates per person). An example of identifying patterns can be if you are interested in seeing the distribution of counties across clusters for only certain socio-economic factors such as “crimes per 100,000 people” or “percentage of population with high school education”). In doing so, a researcher can start to formulate hypotheses as to how the distribution of counties change across cluster types (low, medium, and high performing) as the threshold for the socio-economic variables change (e.g., seeing the distribution change when only including counties with greater than or equal to 50\% of population having high school education vs. 95\%). Lastly, given a hypothesis like this, the map allows you to drill into the data and understand if the hypothesis is valid or not.

\begin{figure*}
\centering
\includegraphics[width=15.5cm]{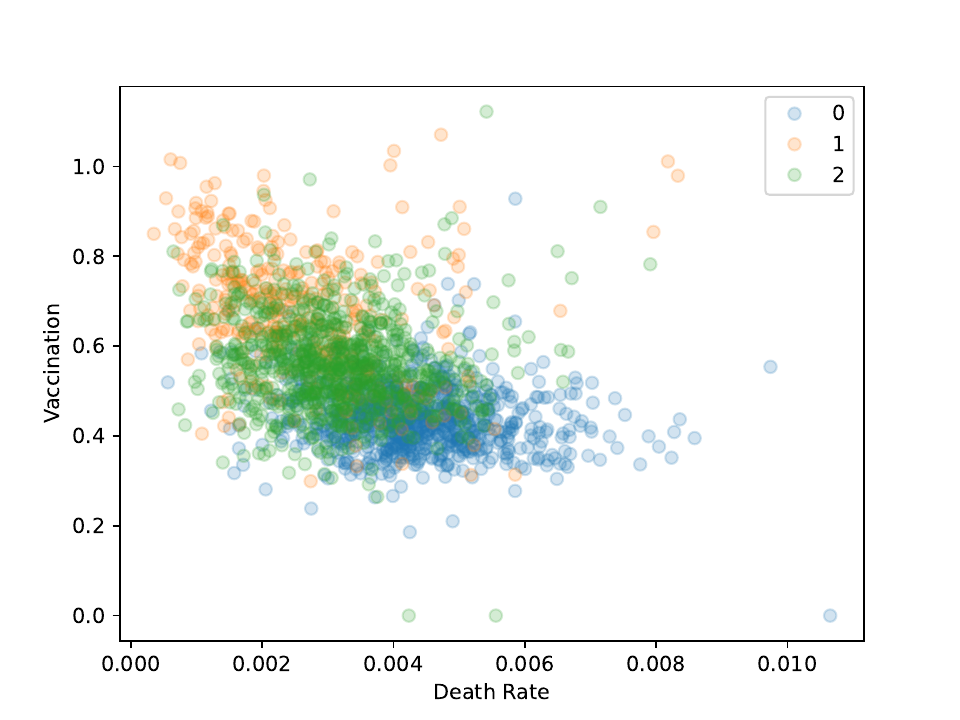}
\caption{2 dimensional comparison of clusters across two variables of Vaccination Rates vs Death Rates}
\end{figure*}

\begin{figure*}
\centering
\includegraphics[width=15.5cm]{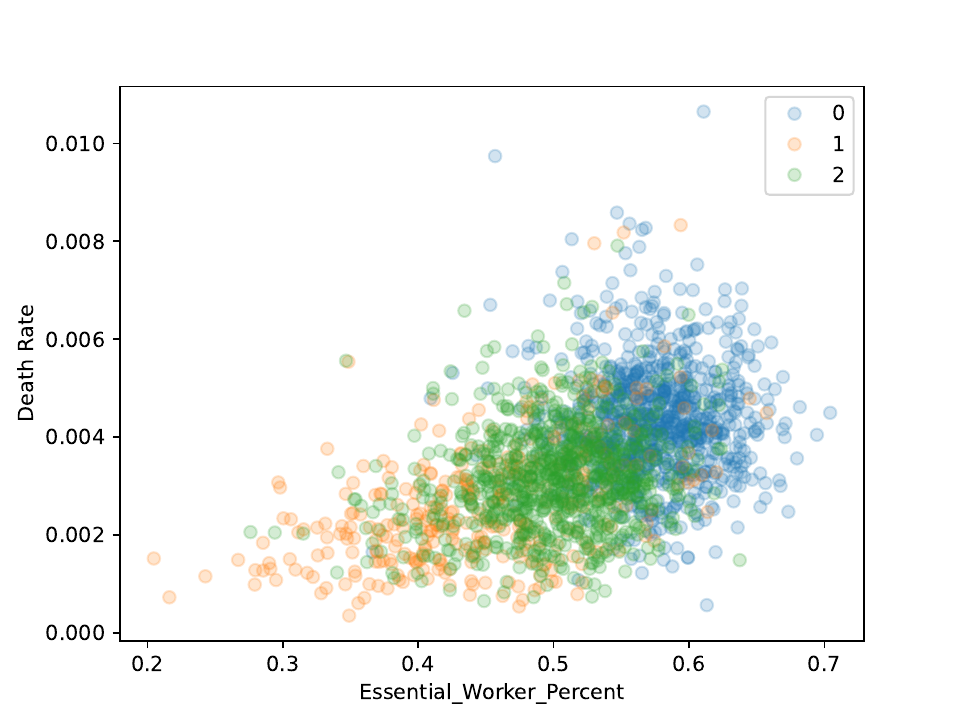}
\caption{2 dimensional comparison of clusters across two variables of Death Rates vs Percentage of Essential Workers}
\end{figure*}

\section{Discussion}

\subsection{Implications of Findings}
The clustering analysis provides crucial insights into the effectiveness of pandemic responses at the county level. Notably, the disparities in healthcare infrastructure and socio-economic conditions play a pivotal role in shaping the outcomes. These findings underscore the need for tailored public health strategies that address the specific needs and challenges of different county profiles.

\subsection{Policy Recommendations}
Based on the results, we recommend:
\begin{itemize}
    \item Enhancing healthcare infrastructure and accessibility in low-performing counties.
    \item Targeted public health campaigns to improve vaccination rates and compliance with health guidelines in medium- and low-performing counties.
    \item Continuous monitoring and data-driven adjustments to pandemic response strategies to mitigate disparities.
\end{itemize}

\subsection{Innovations and Analytical Insights}
Our most significant innovation is our aggregation of multiple datasets to create a detailed repository of information for the application of our unsupervised learning methods. We combined 18 datasets, spanning socioeconomic factors and COVID-specific data, to create a powerful, combined dataset to drive meaningful observations that can inform decision-making in future pandemics to prevent similar societal impact.

Another creative aspect of our approach is to cluster across counties to identify similarities between similar communities across state lines, and then re-focus the analysis to state-specific results. After clustering across 2000 U.S. counties with sufficiently rich datasets, we then looked at state-level distributions of counties. These results were informative and can provide citizens interested in certain state-specific data with insights. Our final meaningful innovation is the creative, interactive visualization techniques we used to leverage our dataset. By summarizing our results in an easily interpretable form across three clusters of counties—with the number of clusters determined via the "elbow diagram" methodology—the team was able to condense our complex dataset into a set of easily interpretable results.

\subsection{Future Directions}
The objective of this study is to identify clusters of counties that vary in performance of combating COVID-19. To do this, we integrated various datasets together at the county level, executed the K-Means clustering algorithm, and analyzed analytically and visually the cluster results. The experiments were set up to test different numbers of K-Means clusters so that the output can be utilized practically as opposed to being so granular with little difference amongst the clusters that the results could not inform future work.

Overall, the study proved to be effective. We were able to identify clear decision boundaries of the county data when utilizing three clusters. Additionally, the results are practical in the sense that when features are analyzed for importance, we can clearly see difference amongst the clusters. Also, when visualizing the clusters and analyzing by the different features, one can explore the data in an effective way that can lead to future research and hypothesis testing.

In conclusion, our analytics combined with the interactive visuals can be utilized to inform how low and medium performing counties can be more like the high-performing counties. This can be done by analyzing the feature values for the two counties to be compared (e.g., Euclidean distance between a lower performing county feature and one higher performing county feature). The counties to analyze can also be selected by visually interacting with the Tableau map; if a user is focused on a specific geography (such as a state), the user can then identify a nearby high-performing county and a nearby lower performing county to begin the analysis to help the lower county improve response to COVID-19 and future pandemics.

Future work should be aimed at piloting improvement opportunities at some of the lower performing counties. This can be done by taking the largest Euclidean distance between any of the variables and identifying initiatives that can potentially help bridge the gap. The results should then be studied to see if the outcomes indeed reduce the impact of COVID-19 or future pandemics. Additionally, other data sets should be included at the county level or at even a lower granularity to inform the study. With more data and ideally with the inclusion of more granular datasets as they are developed, the targeted improvement opportunities can become more targeted and, ideally, more effective.

\bibliographystyle{plainnat}
\bibliography{references}

\end{document}